# A Nonperturbative Study of Quarkonium Systems

J. P. Ma and B. H. J. McKellar

Recearch Center for High Energy Physics

School of Physics

University of Melbourne

Parkville, Victoria 3052

Australia

**Abstract**:

Using Nonrelativistic QCD on the lattice we studied the mass spectrum of quarkonium systems nonperturbatively for a range of the bar quark mass. We determined two products of the matrix elements involved in quarkonium decays and studied the mass dependence of the results. We predict from our calculations the leptonic decay width of $\Upsilon$, and use the mass dependence to predict the leptonic decay width of $J/\psi$. These calculations agree with the experimental results. In lattice NRQCD an additional parameter $n$ is introduced, and we study the sensitivity of our results to the choice of $n$.

1. **Introduction**. In quarkonium the heavy quark moves with a small velocity $v^2$, so nonrelativistic quantum chromodynamics (NRQCD) may be used as a good approximation to describe quarkonium systems. Recent studies have shown that quarkonium systems can be well approximated on lattice by solving NRQCD nonperturbatively[1–4]. This opens an opportunity for precise tests of QCD. It is important to distinguish these *first principles* calculations from model calculations, for example potential model calculations, which may be inspired by QCD, but which depend on phenomenological parameters.

We study quarkonium systems using NRQCD on a $16^3 \times 48$ lattice. We use 20 quenched gauge configurations with $\beta = 6.0$. In previous studies[4], the mass spectrum of bottonium was extensively studied and shown to agree precisely with experiment[4]. We emphasize in our work decay matrix elements, one of which is related to relativistic corrections. However in order to set the physical scale, we need to calculate the mass spectrum. We have simulated quarkonium systems for a range of the heavy quark mass to study the mass dependence of the matrix elements, in the hope that it may be possible to extrapolate our results for the matrix elements from bottonium to charmonium. We must approach charmonium by extrapolation because the lattice spacing at $\beta = 6.0$ is too small for direct simulation of charmonium with lattice NRQCD[5]. Another aspect of our work is to study the $n$-dependence of our results, where $n$ is a parameter introduced into lattice NRQCD for numerical reasons.

The action of NRQCD for quarks on a lattice can be defined as[6]:

$$S_Q = \sum_x \left\{ \psi^\dagger(x)\psi(x) - \psi^\dagger(\mathbf{x}, t+1)(1 - \frac{H_0}{2n})^n U_t^\dagger(\mathbf{x}, t)(1 - \frac{H_0}{2n})^n \psi(\mathbf{x}, t) \right\}. \quad (1)$$

In Eq.(1) $H_0$ is the Hamitonian operator on the lattice, $U_\mu(x)$ is the gauge link, and $n$ is an integer parameter. We only retain the action up to order $v^2$. In our work tadpole improvement[7] is implemented, i.e., each gauge link $U_\mu$ is replaced by $U_\mu/u_0$, where $u_0 = \langle \frac{1}{3}\text{Tr}U_{\text{plaq}} \rangle^{\frac{1}{4}}$. For the configurations used here $u_0 = 0.8778$. With this action the quark propagator $G(\mathbf{x}, t)$ satisfies the evolution equation

$$G(\mathbf{x}, t+1) = \delta_{\mathbf{x},0}\delta_{t+1,0} + (1 - \frac{H_0}{2n})^n U_t^\dagger(\mathbf{x}, t)(1 - \frac{H_0}{2n})^n G(\mathbf{x}, t)$$
$$H_0 = -\frac{\Delta^{(2)}}{2\hat{M}_Q} - h_0, \ h_0 = \frac{3(1-u_0)}{\hat{M}_Q} \tag{2}$$

with $G(\mathbf{x},t)=0$ for $t < 0$. In Eq.(2) $\Delta^{(2)}$ is the lattice Laplacian, and $\hat{M}_Q$ is the quark mass parameter in lattice units. We subtracted a constant in $H_0$, this has an effect that the results from the mean field theory for the mass renormalization constant $Z_m$ and the zero point energy $E_0$ are independent of $n$ and also of $\hat{M}_Q$. The integer $n$ is introduced to avoid the numerical instability when high momentum modes occur[8]. The introduction of $n$ has an effect only at order $a^2$, where $a$ is the lattice spacing. Therefore one expects that results from simulations should not have a strong dependence on $n$ as $a \to 0$. An estimate[6] suggests that $n$ should be larger than $\frac{1.15}{\hat{M}_Q}$ at $\beta = 6.0$ to avoid the instability. With this estimate $n = 1$ should be enough large for $M_Q = 2.0$. However we will see that numerical instability may still occur for $\hat{M}_Q$ around 2.0 and this has effects in the determination of the matrix elements. The numerical values of the matrix elements determined with $n = 1$ and $n = 2$ are significantly different for $\hat{M}_Q \leq 2.0$. For the mass spectrum this $n$-dependence is not significant.

With the action of Eq.(1) spin-symmetry is an exact symmetry, hence quarkonia with the same orbital angular momentum have the same mass. We will only consider $S$-wave and $P$-wave quarkonia without radial excitations.

2.**The Mass Spectrum**. Since the action in Eq.(1) is nonrelativistic, the absolute energy scale is unknown. However the zero-point energy $E_0$ of a quark can be calculated perturbatively. The exact mass of a quarkonium state, for example an $S$-wave state, is related to $E_0$ by:

$$M_S = 2(a^{-1}Z_M \hat{M}_Q - E_0) + E_S. \tag{3}$$

Here $E_S$ is the nonrelativistic energy of the quarkonium and it can be measured in lattice simulations. We will use Eq.(3) to determine the mass of a $S$-wave quarkonium. For $Z_M$ and $E_0$ we use the results from the mean field theory. With $H_0$ given in Eq.(2) they are:

$$\hat{E}_0 = aE_0 = -\ln u_0, \quad Z_M = \frac{1}{u_0} \qquad (4)$$

Although the results for one loop corrections to $E_0$ and $Z_M$ in Eq.(2) exist, we consider it is not necessary to include them because we employ NRQCD in Eq.(1) only at the order of $v^2$ and the corrections with the tadpole improvement are small.

To determine $E_S$ we measure the correlator on the lattice:

$$H_S(t) = \sum_{\mathbf{x}} \langle 0|\chi^\dagger(\mathbf{x},t)\psi((\mathbf{x},t)\psi^\dagger(0)\chi(0)|0\rangle \qquad (5)$$

In Eq. (5), $\chi(x)$ is the field for the anti-quark. For large $t$, $H_S(t)$ takes the asymptotic form:

$$H_S(t) \approx |\langle S|\chi^\dagger\psi|0\rangle|^2 e^{-tE_s}. \qquad (6)$$

From this we determine $\hat{E}_S$ and the matrix element. We measure the correlators with $2^4$ initial points for $n = 1$ and $n = 2$. We used the standard fit to extract $\hat{E}_S$ and the matrix element. Our results for $\hat{M}_S = aM_S$ are given in table 1.

**Table 1**

| $\hat{M}_Q$ | 1.5 | 1.7 | 2.0 | 2.3 | 2.6 |
|---|---|---|---|---|---|
| $\hat{M}_S(n=1)$ | 5.2436(3) | 5.5171(5) | 5.9363(5) | 6.5440(3) | 6.9391(4) |
| $\hat{M}_S(n=2)$ | 5.2132(5) | 5.4351(5) | 5.8552(4) | 6.3488(4) | 6.8828(5) |

From Table 1 one can see that the masses determined from $n = 1$ and $n = 2$ are not significantly different, the differences are only at $1-2\%$ in the $\hat{M}_Q$-range we consider. We also measured the correlator for $P$-wave quarkonium to determine the mass splitting between the $S$-wave and $P$-wave quarkonium at $\hat{M}_Q = 2.6$; the result is

$$\Delta M_{P-S} = 0.352(6), \quad \text{for } n = 1$$
$$\Delta M_{P-S} = 0.331(8), \quad \text{for } n = 2. \qquad (7)$$

Given the fact that the mass spliting is not sensitive to the quark mass in the range considered here, we use the result with $n = 2$ in Eq.(7) and use the experimental result for bottonium to estimate the lattice spacing. We obtain $a^{-1} = 1.33(3)$GeV. In this estimation one should keep in mind that the systematic error is larger than the statistical error quoted here. One source of the systematic error is the neglect of the terms of order $v^4$ in the action. For bottonium $v^2 \approx 0.1$. This means the systematic error due to the $O(v^4)$ terms is about 10% for $\Delta M_{P-S}$, and hence also at least 10% in the estimate of $a$. With this we conclude that the $S$-wave quarkonium simulated here at $\hat{M}_Q = 2.6$ approximately corresponds to bottonium with the predicted mass $M_S = 9.15(20)$GeV. The pole mass of the b-quark can also be determined: $M_b = 3.9(1)$GeV.

Our results for the spectrum are compatible with the results of previous studies by other groups. It is interesting to make a comparson with the results from the precise study of [4], where the terms at the order of $v^4$ are included in the lattice action of NRQCD and some improvement reducing the effect of the finite lattice spacing is also made. The results from there are that the quarkonium at $\hat{M}_Q = 1.71$ corresponds to bottonium, the pole mass of $b$ quark is 4.7GeV and the inverse of the lattice spacing is 2.4GeV. Comparing these with our results above gives a feeling how significant the effect of the accurate action of [4] can be in the physical results. The lattice spacing determined here is with a factor of 2 larger than that of [4]. This is expected since the action employed in [4] is more accurate in $v$ than ours and is improved to reduce the effect of the finite lattice spacing.

**3. The Matrix Elements.** Recently Bodwin, Braaten and Lepage[9] have treated quarkonium systems rigorously within QCD. This contrasts with earlier treatments within the potential model. A series of factorized forms for the decay and production rate of quarkonia were obtained, where the nonperturbative physics is represented through NRQCD matrix elements. With this work one can also systematically account for relativistic corrections in decay and production processes. For example, the decay rate for

$^1S_0 \to \gamma\gamma$ and for $^3S_1 \to \ell^+\ell^-$ can be written as

$$\Gamma(^1S_0 \to \gamma\gamma) = 2\pi Q^4 \alpha^2 \{ \frac{1}{M_Q^2} \langle 0|\chi^\dagger \psi|^1S_0\rangle \langle ^1S_0|\psi^\dagger \chi|0\rangle$$
$$- \frac{4}{3M_Q^4} \text{Re}\{\langle 0|\chi^\dagger (\frac{-i}{2} \overleftrightarrow{\mathbf{D}})^2 \psi|^1S_0\rangle \langle ^1S_0|\psi^\dagger \chi|0\rangle\}\}$$
$$\Gamma(^3S_0 \to \ell^+\ell^-) = \frac{2\pi Q^2 \alpha^2}{3} \{ \frac{1}{M_Q^2} \langle 0|\chi^\dagger \sigma_i \psi|^3S_1\rangle \langle ^3S_1|\psi^\dagger \sigma_i \chi|0\rangle$$
$$- \frac{4}{3M_Q^4} \text{Re}\{\langle 0|\chi^\dagger \sigma_i (\frac{-i}{2} \overleftrightarrow{\mathbf{D}})^2 \psi|^3S_1\rangle \langle ^3S_1|\psi^\dagger \sigma_i \chi|0\rangle\}\} \quad (8)$$

at the leading order in $\alpha$ and $\alpha_s$. In Eq.(8) terms with $M_Q^{-4}$ are relativistic corrections of order $v^2$. The matrix elements in Eq.(8) are defined in NRQCD and they can only be calculated nonperturbatively. For bottonium the state $^1S_0$ has still not been found experimentally. Since the spin-symmetry is an exact symmetry in the approximation used here, the two products of the matrix elements in $\Gamma(^1S_0 \to \gamma\gamma)$ are equal to those in $\Gamma(^3S_0 \to \ell^+\ell^-)$. They have also a simple relation to the matrix elements in hadronic decay. We study these products directly, introducing the notation:

$$F_1 = \langle 0|\chi^\dagger \psi|^1S_0\rangle \langle ^1S_0|\psi^\dagger \chi|0\rangle$$
$$G_1 = \text{Re}\{\langle 0|\chi^\dagger (\frac{-i}{2} \overleftrightarrow{\mathbf{D}})^2 \psi|^1S_0\rangle \langle ^1S_0|\psi^\dagger \chi|0\rangle\}. \quad (9)$$

The quantity $F_1$ has dimension 3 in mass and $G_1$ has dimension 5 in mass. $F_1$ is proportional to the square of the absolute value of the wave function at the origin. On the lattice $F_1$ can be extracted form the correlation function in Eq.(5). To measure $G_1$ we construct a suitable correlation function, in which we use the covariant centered difference on the lattice for the covariant derivative $\mathbf{D}$. The quantity $G_1$ is also studied in [10], where the same lattice action is used as in Eq.(1). However, it is claimed in [10] that the quarkonium simulated at $\hat{M}_Q = 1.5$ corresponds to bottonium, which is in contrary to our result above. Our results for $\hat{F}_1 = a^3 F_1$ and $\hat{G}_1 = a^5 G_1$ are given in Table 2.

**Table 2**

| $\hat{M}_Q$ | 1.5 | 1.7 | 2.0 | 2.3 | 2.6 | 2.9 |
|---|---|---|---|---|---|---|
| $\hat{F}_1(n=1)$ | 2.86(4) | 1.92(4) | 1.58(2) | 1.56(2) | 1.68(2) | 1.88(2) |
| $\hat{F}_1(n=2)$ | 1.16(2) | 1.20(1) | 1.35(1) | 1.66(2) | 1.82(2) | 2.10(3) |
| $\hat{G}_1(n=1)$ | 1.19(1) | 0.64(1) | 0.45(1) | 0.43(1) | 0.48(1) | 0.58(2) |
| $\hat{G}_1(n=2)$ | 0.224(8) | 0.248(6) | 0.31(1) | 0.41(1) | 0.53(1) | 0.67(2) |

We have seen that the mass spectrum is not sensitive to the parameter $n$, however the results for $\hat{F}_1$ and $\hat{G}_1$ are quite different for different $n$. The parameter $\hat{F}_1$ determined with $n=1$ decreases linearly with $\hat{M}_Q$ from 2.9 to 2.3, then it increases as $\hat{M}_Q$ decreases further. From physical arguments one expects that $F_1$ decreases if $M_Q$ decreases. The possible reason for the deviation from these expectation can be that $n=1$ is still not large enough to prevent the numerical instability caused by high momentum modes arround $\hat{M}_Q = 2.0$ in the determination of $\hat{F}_1$. We conclude that the calculation with $n=1$ can not give correct results for $\hat{M}_Q \leq 2.0$, where local sources are used to calculate quark propagators in Eq.(2). In the following we will only take the results with $n=2$ for discussions.

From our data $\hat{F}_1$ is approximately proportional to $\hat{M}_Q$ in the $\hat{M}_Q$ range from 1.7 to 2.9: $\hat{F}_1 \approx 0.7 \hat{M}_Q$. This behaviour is expected, since $F_1$ is expected to be proportinal to $M_Q$. In our approximation $M_Q = a^{-1} Z_M \hat{M}_Q$ is proportional to $\hat{M}_Q$. But at $\hat{M}_Q = 1.5$, $\hat{F}_1$ deviates from this proportionality relation, as can be seen in Fig.1 where the straight line represents the relation $\hat{F}_1 = 0.7 \hat{M}_Q$. Similarly, we find that the values of $\hat{G}_1$ for $\hat{M}_Q$ from 2.0 to 2.9 satify the relation $\hat{G}_1 = 0.078 \hat{M}_Q^2$ very well, but the values at $\hat{M}_Q = 1.5$ and 1.7 deviate from the relation. We suggest that the reason for the deviations at lower $\hat{M}_Q$ is the same as discussed above. In Fig.2 we plot the results for $\hat{G}_1$, the curve is $\hat{G}_1 = 0.078 \hat{M}_Q^2$. From Fig.1 and Fig.2 one observes that $\hat{G}_1$ for $\hat{M}_Q$ from 2.0 to 2.9 fits a parabola much better than $\hat{F}_1$ fits a straight line. This is confirmed by the $\chi^2$ of the fits: for $\hat{G}_1$ $\chi^2$ is 0.61, while $\chi^2$ of the fit for $\hat{F}_1$ is 20 for $\hat{M}_Q$ from 1.7, 2.0 and 2.3 to 2.9. If we include the data point at $\hat{M}_Q = 1.5$ into the fit, $\chi^2$ becomes 45!

With these results one can predict the leptonic decay width of $\Upsilon$. We use the following formula for doing this:

$$\Gamma(^3S_1 \to \ell^+\ell^-) = 2\pi Q^2 \alpha^2 \left\{ (1 - \frac{16}{3\pi}\alpha_s(M_Q))\frac{1}{M_Q^2}(a^{-1})^3 \hat{F}_1 - \frac{4}{3}\frac{1}{M_Q^4}(a^{-1})^5 \hat{G}_1 \right\}. \quad (10)$$

Here the one loop correction from QCD in the coffecient of $F_1$ term is also included. For $\Upsilon$ we take $M_Q = M_b = 4.7\text{GeV}$, $\alpha = 1/128$ and $\alpha_s(M_b) = 0.20$, where we used the one-loop $\beta$-function and the experimental value for $\alpha_s(M_Z) = 0.115$ to evolve $\alpha_s$ to $\mu = M_b$. We obtain $\Gamma(\Upsilon \to e^+e^-) \approx 1.7\text{KeV}$. Our result is in agreement with the experimental result, $\Gamma(\Upsilon \to e^+e^-) \approx 1.3\text{KeV}$. However, as discussed for the $a^{-1}$ determination before, a large systematic error can occur in our result and it can not be taken as a precise prediction. That is why we have not quoted any error in our estimate above. The $G_1$ term gives in Eq.(10) a small negative contribution which is only 5% of the decay width. Another way to use our lattice results to predict the decay width is to extract directly from lattice simulations the dimensionless quantities $F_1/M_Q^3 = \hat{F}_1/(Z_M \hat{M}_Q)^3$ and $G_1/M_Q^5 = \hat{G}_1/(Z_M \hat{M}_Q)^5$ for the estimate of the decay width. The advantage is that a direct use of $a^{-1}$ is avoided. But since these quantities depends on a power of the quark mass, a precise location of $\Upsilon$ in the parameter space of $\hat{M}_Q$ is needed for a precise prediction. The decay width determined in that way with $\hat{M}_b = 2.6$ is 2.8KeV, which is not too far away from either the experimental value or our other estimate.

With the mass dependence of $F_1$ and $G_1$ we can also estimate the leptonic decay width of $J/\psi$. Assuming $M_b : M_c = \hat{M}_b : \hat{M}_c$ and taking $M_c = 1.3\text{GeV}$ we obtain $\hat{M}_c = 0.719$. For $\alpha_s$ at $\mu = M_c$ we use the experimental value $\alpha_s(M_\tau) = 0.355$ and evolve this value to $\mu = M_c$. Using $\hat{G}_1 = 0.078 \hat{M}_Q^2$ and $\hat{F}_1 = 0.7 \hat{M}_Q$ we obtain with Eq.(10): $\Gamma(J/\psi \to e^+e^-) \approx 7.1\text{KeV}$, which is not too far from the experimental result: $\Gamma(J/\psi \to e^+e^-) \approx 5.4\text{KeV}$. In our estimates the QCD corrections are very important, in particular the factor $(1 - \frac{16}{3\pi}\alpha_s)$ is 0.29 for $J/\psi$, and 0.66 for $\Upsilon$, but without the QCD correction the factor is 1. We also see that the relativistic correction is large for $\hat{G}_1$ because of the relatively small $c$ quark mass. This correction is at the level of 20%. Our final result is the determination of $\Gamma(\eta_c \to \gamma\gamma)$. With the extrapolated values of $\hat{F}_1$ and $\hat{G}_1$ we obtain: $\Gamma(\eta_c \to \gamma\gamma) \approx$ 23KeV, where the one loop QCD correction is included. Comparing the experimental result $\Gamma(\eta_c \to \gamma\gamma) \approx 6.6$KeV our value is three times too large. The fact that our prediction for $\Gamma(J/\psi \to e^+e^-)$ is close to the experimental result gives some surport to the validity of the mass dependence for $\hat{F}_1$ and $\hat{G}_1$. However, the discrepancy in $\Gamma(\eta_c \to \gamma\gamma)$ indicates that the mass dependence may need to be modified in order to get both $\Gamma(\eta_c \to \gamma\gamma)$ and $\Gamma(J/\psi \to e^+e^-)$ in agreement with experiment. In our work we neglected the $O(v^4)$ terms in the action and have taken only the tree-level results for renormalization constants to convert lattice results into those of the continuum. One can hope that including the $O(v^4)$ terms, and by taking higher order effects in renormalization constants into account, the results will be improved. The effect of renormalization constants may be significant in converting lattice results for $F_1$ to continuum results, since at the one loop level, the operator in $F_1$ is mixed with that in $G_1$. All of the relevant renormalization constants are not yet available at the one-loop level.

**4. Summary.** In this work we have studied the properties of quarkonium systems by solving NRQCD on lattice nonperturbatively. The results for the mass spectrum are compatible with the results from other groups and with experiment. We studied two products of matrix elements involved in quarkonium decays, one of which is related to the relativistic corrections. Our prediction of $\Gamma(\Upsilon \to e^+e^-) \approx 1.7$KeV is in agreement with experiment. It is found that the relativistic correction is small (5%) as expected. It should be stressed that this result is a QCD prediction and is not a result from model calculations. We simulated quarkniom systems in a range of the quark mass and found the mass dependence of $\hat{F}_1$ and $\hat{G}_1$. $\hat{F}_1$ is proportional to $\hat{M}_Q$ as expected and $\hat{G}_1$ is proportional to $\hat{M}_Q^2$. With these relations we extropolate our prediction for $\Gamma(\Upsilon \to e^+e^-)$ to $\Gamma(J/\psi \to e^+e^-)$. The prediction for $\Gamma(J/\psi \to e^+e^-)$ is in reasonable agreement with experiment. A large relativistic correction ($\approx 22\%$) in charmonium is found. However our prediction for $\Gamma(\eta_c \to \gamma\gamma)$ is 3 times larger than the experimental value.

In NRQCD on lattice a extra parameter $n$ is introduced for preventing numerical instabilities from high momentum modes. We studied the $n$-dependence of our results by taking $n = 1$ and $n = 2$. The mass spectrum does not depend on $n$ significantly. However, the difference between the results for $\hat{F}_1$ and $\hat{G}_1$ with defferent $n$ is large at $\hat{M}_Q \leq 2.0$. In addition, the value of $\hat{F}_1$ determined with $n = 1$ at $\hat{M}_Q \leq 2.0$ increases with decreasing quark mass, this is against physical expectations. We think that the results with $n = 2$ for $\hat{M}_Q$ from 2.0 to 2.9 are reasonable. Our prediction for the decay widths is based on these results.

**Acknowlegment:** The gluon configurations used here are from UKQCD. We would like to thank UKQCD for providing their configurations to us. Discussions with Dr. L. Hollenberg and Dr. T.D. Kieu are acknowleged. The whole calculation was performed on several UNIX work-stations available to us, we thank Dr. M. Munro, the system manager, for helping. This work is supported by Australian Research Council.

# Reference

[1] B.A. Thacker & G.P. Lepage, Phys. Rev. D43 (1992) 196

[2] C.T.H. Davies & B.A. Thacker, Nucl. Phys. B405 (1993) 593

[3] S.M. Catterall et. al. (UKQCD Collabration), Phys. Lett. B300 (1993) 393

[4] C.T.H. Davies et. al. (NRQCD Collabration), Phys. Rev. D50 (1994) 6963, Phys. Rev. Lett. **73** (1994) 2654

[5] C. Morningstar, Phys. Rev. D50 (1994) 5902

[6] G.P. Lepage, L. Magnea & C. Nakhleh, Phys. Rev. D46 (1992) 4052

[7] G.P. Lepage & P. Mackenzie, Phys. Rev. D48 (1993) 2250

[8] C.T.H. Davies & B.A. Thacker, Phys. Rev. D45 (1992) 915

[9] G.T. Bodwin, E. Braaten & L.P. Lepage, Phys. Rev. D51 (1995) 1125

[10] G.T. Bodwin, S. Kim & D.K. Sinclair, Argonne Preprint ANL-HEP-CP-94-91

Figure Caption

Fig.1 The values of $\hat{F}_1$ vs $\hat{M}_Q$. The x-axis is for $\hat{M}_Q$, the y-axis is for $\hat{F}_1$. The points are the data points.

Fig.2 The values of $\hat{G}_1$ vs $\hat{M}_Q$. The x-axis is for $\hat{M}_Q$, the y-axis is for $\hat{G}_1$. The points are the data points.

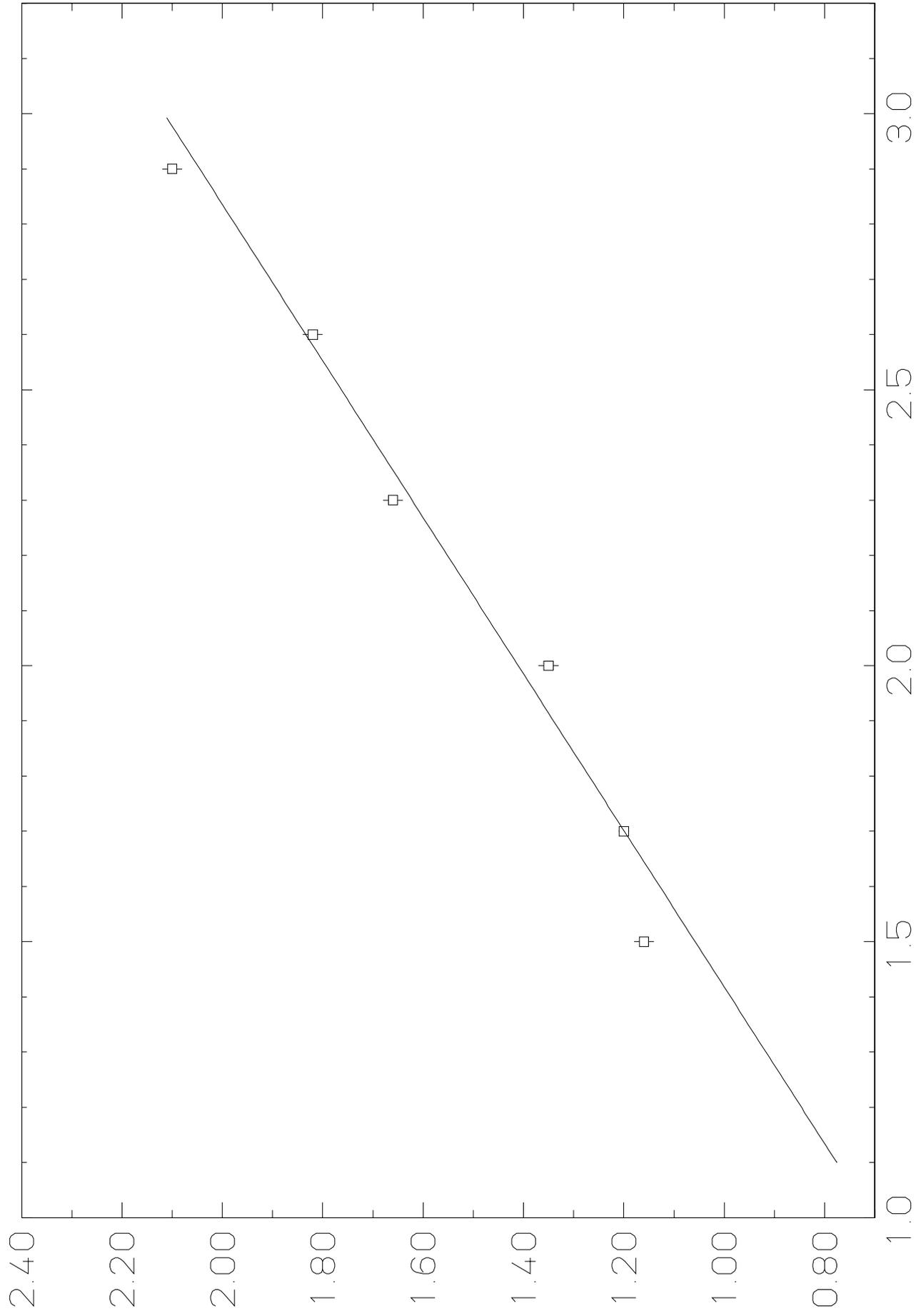

Fig.1

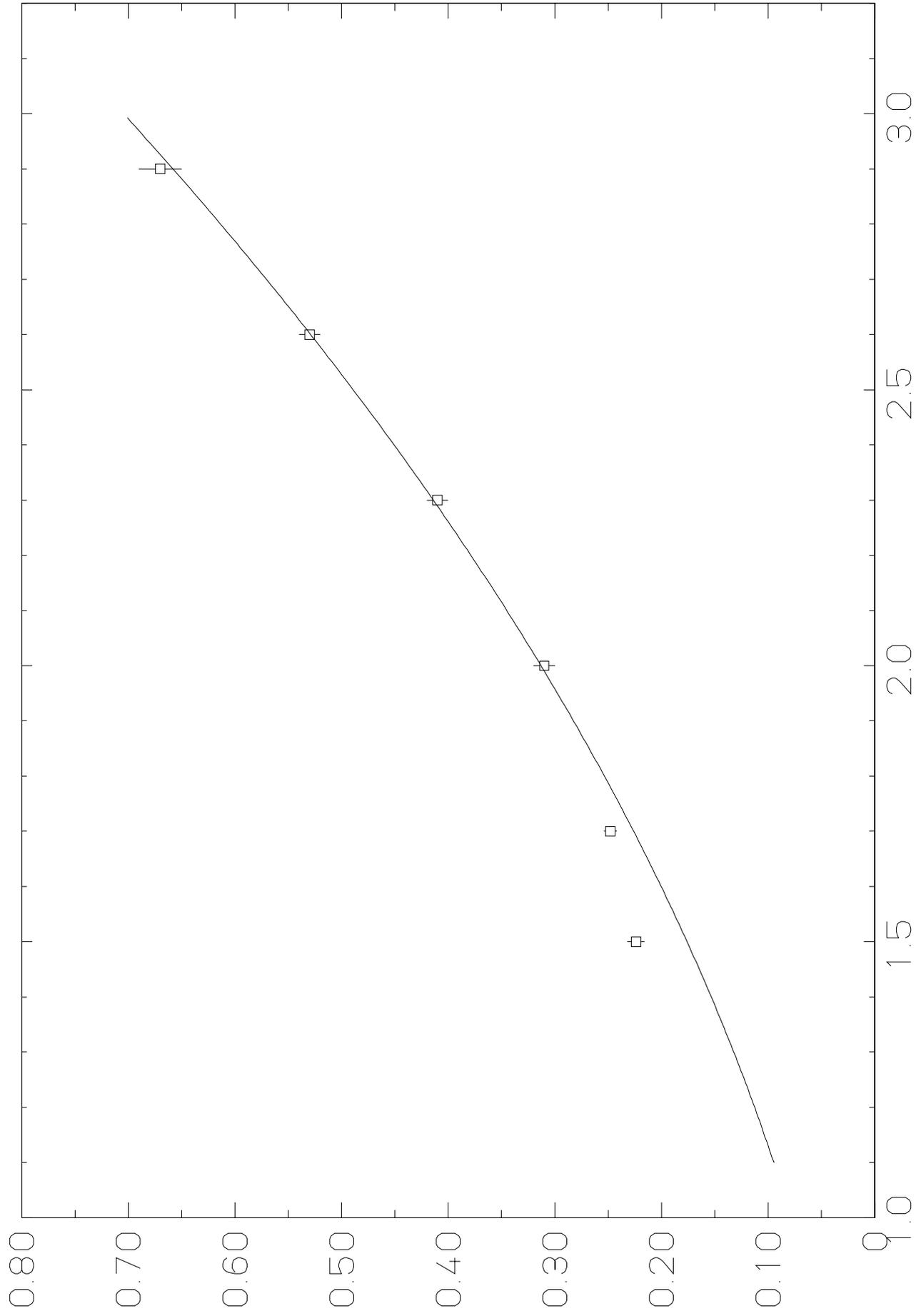

Fig.2